# Nonlinear dynamical systems and bistability in linearly forced isotropic turbulence

Zheng Ran[1], Xing-jie Yuan[1]
[1]Shanghai Institute of Applied Mathematics and Mechanics,
 Shanghai University, Shanghai 200072, P.R.China

**Abstract**  In this letter, we present an extensive study of the linearly forced isotropic turbulence. By using analytical method, we identify two parametric choices, of which they seem to be new as far as our knowledge goes. We prove that the underlying nonlinear dynamical system for linearly forced isotropic turbulence is the general case of a cubic Lienard equation with linear damping. We also discuss a Fokker-Planck approach to this new dynamical system,which is bistable and exhibits two asymmetric and asymptotically stable stationary probability densities.
 .

**Keywords** Isotropic turbulence•Nonlinear dynamical system• Karman-Howarth equation

One of the main goals in the development of the theory of dynamical system has been to make progress in understanding the statistical behavior of turbulence. The attempts to relate turbulence to nonlinear dynamical system motion has received strong impetus from the celebrated papers by Landau [1], Lorenz [2], and Ruelle–Takens [3] and others. Considerable success has been achieved mainly at the low-dimensional model. For fully developed turbulence, many questions remain unanswered [4]. The possibility of a dynamical system approach allows one to capture the fundamental physical mechanisms, such as the energy cascade in three-dimensional turbulence. In particular, one can build a bridge between the traditional description in statistical terms and the dynamical behavior in phase space. For fluids under severe constraints, i.e. the experiments on Rayleigh-Benard convection in small cells, the success was undisputed. For the understanding of fully developed isotropic turbulence, however, the success has been more limited [5]. We observe that the present stage of development of the theory of dynamical systems there has been little quantitative impact on the understanding of high Reynolds number flow [6].

We outline here a new method to obtain a natural second order nonlinear dynamical system based on the analytical solution of linearly forced Karman-Howarth equation. Maintaining a turbulent flow in a more or less stationary state, for better statistics in experiment or convenience in theoretical considerations, requires forcing the flow. There have been developed various kinds of forcing schemes. A special forcing scheme [7] is proposed  in that we may simplify the deterministic models to the bare minimum, in some sense, assuming that usually velocity dependent force term is merely proportional to the velocity field for all positions, and all times. The 'linear forcing' scheme was further studied by several groups [8,9]. We will consider this kind of forced isotropic turbulence governed by the incompressible Navier-Stokes equations

$$\frac{\partial \vec{u}}{\partial t} + (\vec{u} \cdot \nabla)\vec{u} = -\frac{1}{\rho}\nabla p + \nu \nabla^2 \vec{u} + \phi \vec{u}, \qquad (1)$$

Where the incompressibility condition reads

$$\nabla \cdot \vec{u} = 0, \quad (2)$$

where $\phi$ is a positive constant with dimensions of inverse time.

The two-point double and triple longitudinal velocity correlation denoted by $f(r,t)$ and $h(r,t)$ respectively, are defined in a standard way [10]. The Karman-Howarth equation derived from it under the conditions of homogeneity and isotropy reads

$$\frac{\partial}{\partial t}(bf) + 2b^{\frac{3}{2}}\left(\frac{\partial h}{\partial r} + \frac{4h}{r}\right) = 2\nu b\left(\frac{\partial^2 f}{\partial r^2} + \frac{4}{r}\frac{\partial f}{\partial r}\right) + 2\phi bf \quad (3)$$

where $(r,t)$ is the spatial and time coordinates, $\nu$ is the kinematics viscosity, and $b = <u^2>$ denotes the turbulence intensity. Following von Karman and Howarth, we introduce the new variables

$$\xi = \frac{r}{l(t)} \quad (4)$$

where $l = l(t)$ is a uniquely specified similarity length scale. As already noted in the free decay isotropic turbulence case [11,12,13,14,15], equation (3) could be reduced into the following systems:

[1] Two-point double and triple longitudinal velocity correlation, denoted by $f(r,t)$ may be written

$$\frac{d^2 f}{d\xi^2} + \left(\frac{4}{\xi} + \frac{a_1}{2}\xi\right)\frac{df}{d\xi} + \frac{a_2}{2}f = 0 \quad (5)$$

With boundary conditions $f(0) = 1$, $f(\infty) = 0$, and $a_1$ and $a_2$ are constant coefficients.

[2] The turbulent scales equations read

$$\frac{dl}{dt} = a_1 \cdot \frac{\nu}{l} + 2I_1 \cdot \sqrt{b} \quad (6)$$

$$\frac{db}{dt} = -a_2 \cdot \frac{\nu b}{l^2} + 2\phi b + 2I_2 \cdot \frac{b^{\frac{3}{2}}}{l} \quad (7)$$

Where $I_1$ and $I_2$ are two constants of integration.

[3] For third order of correlation coefficient, from the system of Karman-Howarth equations one can derive the following equation

$$\frac{dh}{d\xi} + \frac{4}{\xi}h = -\frac{l}{2b^{\frac{3}{2}}}\frac{db}{dt}f + \frac{1}{2\sqrt{b}}\frac{dl}{dt}\xi\frac{df}{d\xi} + \frac{\nu}{\sqrt{b}l}\left(\frac{d^2 f}{d\xi^2} + \frac{4}{\xi}\frac{df}{d\xi}\right) + \frac{\phi l}{\sqrt{b}}f \quad (8)$$

In fact, the problem is thus reduce to solving a solvable system, furthermore, one can prove that there is a self-closed second order nonlinear dynamical system for $z = z(t) \equiv l^{-2}$:

$$\frac{d^2 z}{d\tau^2} + [\phi - Az] \cdot \frac{dz}{d\tau} + Bz^3 + Cz^2 = 0 \quad (9)$$

Where

$$\tau = -t \quad (10)$$

The above-written system thus depend on the three constant parameters $(a_1, \sigma, \nu)$, where $a_1 > 0$, and [14,15]

$$\sigma = \frac{a_2}{2a_1} \quad (11)$$

Consequently, the parameters could be found

$$A = (a_1 \nu) \cdot [7 + \sigma] \quad (12)$$

$$B = 2(a_1 \nu)^2 \cdot [3 + \sigma] \quad (13)$$

$$C = -2\phi \cdot (a_1 \nu) \quad (14)$$

In order to using the standard mathematical results on nonlinear dynamical system [16], the transformation is performed by the formulae

$$z = \frac{1}{\sqrt{B}} \cdot x - \frac{C}{3B} \quad (15)$$

In can be easily proven that this transformations, in spite of their nonlinear dynamical character, the new nonlinear dynamical system may be found in a standard manner as:

$$\frac{dx}{dt} = y \quad (16)$$

$$\frac{dy}{dt} = -x^3 + \mu_2 x + \mu_1 + y \cdot [\bar{b} x + \bar{\nu}] \quad (17)$$

here:

$$\mu_2 = \frac{2}{3} \cdot \frac{\phi^2}{3 + \sigma} \quad (18)$$

$$\mu_1 = \frac{4\sqrt{2}}{27} \cdot \frac{\phi^3}{[3 + \sigma]^{\frac{3}{2}}} \quad (19)$$

$$\bar{\nu} = -\phi \left\{ 1 + \frac{7 + \sigma}{3(3 + \sigma)} \right\} \quad (20)$$

$$\bar{b} = \frac{7+\sigma}{\sqrt{2(3+\sigma)}} \qquad (21)$$

Reference 16 presents an extensive qualitative study of the phase portraits of cubic Lienard equations with linear damping. There are three cases in the study of this system which are topologically distinct. For linearly forced isotropic turbulence, it is shown that the system belongs to the elliptic $(\bar{b} \geq 2\sqrt{2})$, and the above system has at most one periodic orbit, which is a hyperbolic limit cycle.

We are concerned with the bifurcation solutions of the above system. Firstly, we noticed that the vector field has either one or three singular points, depending on the values of the parameters $\mu_1, \mu_2$. It should be clear that $S_1 \equiv (x, y) = \left( \frac{2\sqrt{2}}{3} \cdot \frac{\phi}{\sqrt{\sigma+3}}, 0 \right)$ is a hyperbolic fixed point of the system. Meanwhile, $S_2 \equiv (x, y) = \left( -\frac{\sqrt{2}}{3} \cdot \frac{\phi}{\sqrt{\sigma+3}}, 0 \right)$ is a non-hyperbolic fixed point.

The former offer services as a candidate for the stationary state of turbulence. For $S_2$, we study the structral stability inheret in this non-hyperbolic fixed point. A center manifold theory for this turbulent dynamical systems has been developed by us. Thus, since the invariant center manifold exists in a sufficiently small neighborhood in $S_2$, all bifurcating solutions will be contained in the lower dimensional center manifold. The vector field restricted to the center manifold is given by

$$\frac{d\xi}{ds} = -\frac{\partial V}{\partial \xi} \qquad (22)$$

Which indicated that the origin is stable. The above differential equation is a derterministic equation. We now want to discuss a stochastic force occurs. The Langevin equation now is

$$\frac{d\xi}{ds} = -\frac{\partial V}{\partial \xi} + \Gamma(s) \qquad (23)$$

where $\Gamma(s)$ is the Gaussian noise

$$\langle \Gamma(s) \rangle = 0, \qquad (24)$$

$$\langle \Gamma(s)\Gamma(s') \rangle = 2Q\delta(s-s'). \qquad (25)$$

$Q$ describes the strength of the fluctuating force. The general Fokker-Planck equation for one variable $\xi$ has the form

$$\frac{\partial P}{\partial s}+\frac{\partial}{\partial \xi}\left[-\frac{\partial V}{\partial \xi}\cdot P\right]-Q\frac{\partial^2 P}{\partial \xi^2}=0. \tag{26}$$

where $P=P(\xi,t)$ is the distribution function.

We note that, many nonequilbrium systems can be described in the vicinity of phase transition points by means of a few relevant variables [17] and , in particular, by means of a few order parameters[18]. As ponited out by Haken , it is helpful to regard order prameters of nonequilibrium systems as counterparts to the order parameters of the Landau theory [19]. They are our further works. Here, we only present some numerical results of the Fokker-Planck equation with the double well potential given in the appendix. The transient distribution functions for different parameters are given in Figure.1. The system is bistable and exhibits two asymmetric and asymptotically stable stationary probability densities.

The aim of this letter was to show that there are natural nonlinear dynamical systems that it has many distinct properties compared with the well known low dimensional dynamical model ( such as Lorenz model ), but that can still have turbulent states and for which many concepts developed in the theory of dynamical systems can be successfully applied. In this connection we advocate a broader use of the universal properties of a wide range of isotropic turbulence phenomena .

The work was supported by the National Natural Science Foundation of China (Grant Nos.11172162, 10572083).

SSSR **42**,116 (1944).

[12] L.I.Sedov, Similarity and dimensional methods in mechanics. Translated from the Russian by V..I. Kisin. (Mir Publishers,1982).

[13] A.I.Korneyev and L.I. Sedov, Theory of isotropic turbulence and its comparison with experimental data. Fluid Mechanics-Soviet Research 5, 37 (1976).

[14] Z.Ran, New Sedov-type solution of isotropic turbulence. Chin. Phys. Lett. 12,4318 (2008).

[15] Z.Ran, One exactly soluble model in isotropic turbulence. Advances and Applications in Fluid Mechanics. 5, 41 (2009).

[16] F.Dumortier and C.Roussea, Cubic Lienard equations with linear damping. Nonlinearity F. 3,1015-1039 (1990).

[17] M.C.Cross and P.C.Hohenberg, Pattern formation outside of equilibrium. Rev.Mod. Phys., 65:851-1112 (1993).

[18] H.Haken, Cooperative phenomena in systems far from thermal equilibrium and in nonphysical systems. Rev. Mod. Phys., 47:67-121 (1975).

[19] L.D.Landau and E.M.Lifshitz, Statistical Physics. Pergramon Press, London (1958).

**APPENDIX.   The double well potential for linearly forced isotropic turbulence**

$$V = \xi^4 + \lambda \xi^2 + \mu \xi, \tag{A.1}$$

where

$$x_s = -\frac{\sqrt{2}}{3} \cdot \frac{\phi}{\sqrt{\sigma+3}}, \tag{A.2}$$

$$\bar{b} = \frac{\sigma+7}{\sqrt{2(\sigma+3)}}, \tag{A.3}$$

$$\delta = \left[1 + \frac{2(\sigma+7)}{3(\sigma+3)}\right] \cdot \phi, \tag{A.4}$$

$$C_2 = -\frac{3x_s}{\delta}, \tag{A.5}$$

$$C_3 = -\frac{1}{\delta} - \frac{3x_s}{\delta^2}\bar{b} - \frac{18x_s^2}{\delta^3}, \tag{A.6}$$

$$\lambda = -\frac{3}{8}\left[-\frac{C_2}{3C_3}\right]^2, \tag{A.7}$$

$$\mu = -\left[-\frac{C_2}{3C_3}\right]^3. \tag{A.8}$$

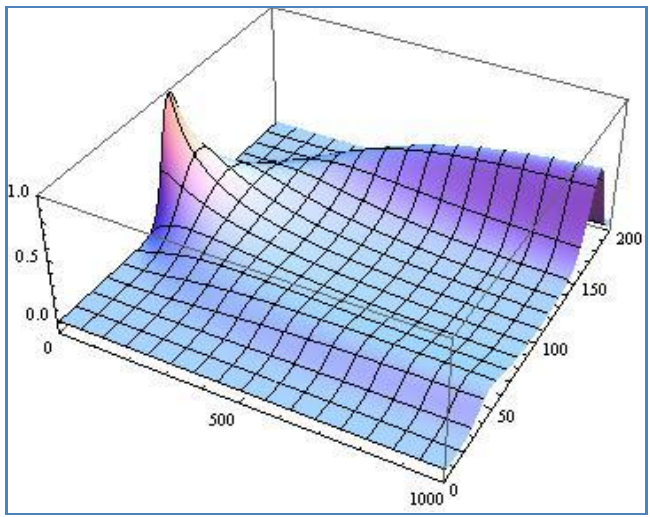

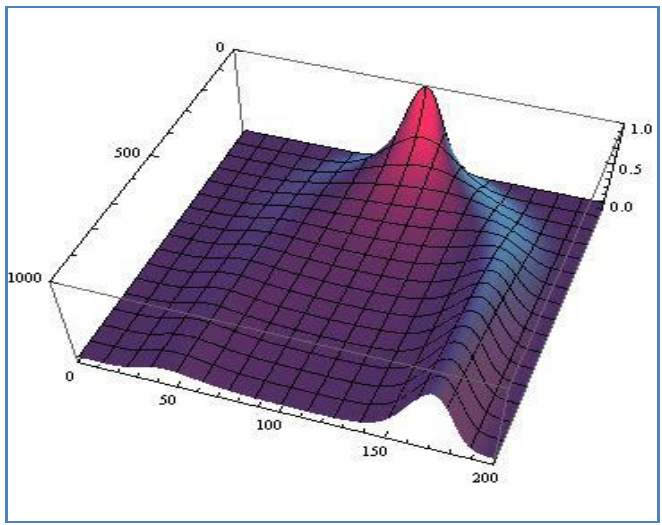

Figure.1  The transient distribution functions for different parameters

$\lambda = -\dfrac{1}{8}$

$\mu = -0.0125$

$Q = 0.01$